\documentclass[prl,twocolumn,epsf,psfig]{revtex4}
\usepackage{graphicx}
\DeclareGraphicsExtensions{.pdf,.png,.gif,.jpg}
\usepackage{float}
\usepackage{subfig}

\begin{document}

\title{Polarization-induced phase separation and re-entrant transition of two component-fermions in a one-dimensional lattice}

\author{Theja N. De Silva}
\affiliation{Department of Chemistry and Physics,
Georgia Regents University, Augusta, GA 30912, USA.}
\begin{abstract}

By investigating the compressibility of one-dimensional lattice fermions at various filling factors, we study the phase separation and re-entrant transition within the framework of the Bethe ansatz method. We model the system using the repulsive Hubbard model and calculate compressibility as a function of polarization for arbitrary values of chemical potential, temperature, and interaction strength. For filling factors $ 0 < n < 1$, compressibility is a non-monotonic function of polarization at all thermodynamic parameters. The compressibility reveals a phase transition into a phase-separated state for both low and intermediate temperatures at intermediate interactions as one increases the polarization. For certain filling factors, we find the re-entrant transition into the mixed phase at a higher polarization.
\end{abstract}

\maketitle

\section{I. Introduction}

Ultracold fermionic atoms in optical lattices are currently attracting a great deal of interest due to the possibility of impressive experimental simulation of rich physics associated with the strongly correlated condensed matter systems~\cite{opl1, opl2}. While Feshbach resonance and laser intensities provide unprecedented control of atom-atom interactions in the optical lattice, laser interference phenomena provides the control of dimensionality. Two hyperfine states of Fermi atoms play the role of the up and down spins
 of the electrons. Unlike condensed matter systems, the population of spin up and down particles can be independently controlled by using a radio-frequency field~\cite{oplE1, oplE2, oplE3, oplE4, oplE5, oplE6, oplE7, oplE8, oplE9, oplE10}. As a result, polarization can be maintained at any desired value between 0 and 100\%. The term polarization here refers to a density imbalance of two hyperfine states.

This study was motivated by the possibility of using ultra-cold atoms to engineer condensed matter systems. Condensed matter systems, such as transition metal oxides and rare-earth materials show collective and interaction dominated phenomena due to the electron-electron correlation effects. These phenomena such as the Mott-insulator transition and magnetism in strongly correlated materials is believed to be explained by the Hubbard model. The repulsive Hubbard model is the simplest model capable of explaining both metallic and insulating like behaviors, as well as the magnetic properties caused by electron correlation. The insulating states can be either a band insulator, caused by the Pauli exclusion of fermions, or a Mott-insulator, caused by strong on-site interactions. In the strong interaction limit, the localized magnetic properties depend on various parameters such as filling factors, orbital occupations, crystal field effects, and Hund's coupling strength.

Over the years, the Hubbard model has been the center of intense research as it captures the behavior of parent superconducting compounds and other magnetic materials. The studies of phase separation for the Hubbard model intensified after the experimental indication of phase separation of hole-rich and hole-poor regions in cuprate superconducting materials~\cite{psCU}. Strongly correlated electrons and holes are expected to play a key role in these materials and their phase separation is believed to hinder the superconductivity. If the Hubbard model is the correct model for the parent compound of superconductors, can it be used to explain phase separation? This is the question that inspired us to study the phase separation of an exactly solvable one-dimensional model relevant to a flexible cold-atom experiment. It has been theoretically shown that there is no phase separation for two-dimensional bipartite lattices at any filling factors at finite temperatures~\cite{tdps}. In contrast, the phase separation for the one dimensional Hubbard model is confirmed only close to a half filling in the presence of a critical magnetic field~\cite{odps}. Finite-temperature phase separation for the one dimensional Hubbard model away from half filling has not been intensively investigated except for special cases~\cite{bosT, LB}.

In this paper we investigate the phase separation of one dimensional lattice fermions by Bethe ansatz (TBA) numerical method. We use the one dimensional Hubbard model as an effective model to describe the population-imbalanced two-hyperfine mixture in the optical lattice. We study phase separation by calculating the compressibility for various parameter regimes. By investigating the compressibility, we find phase separation at finite temperatures and at intermediate interactions as one increases the polarization. For some filling factors, we find a re-entrant transition into a mixed phase at a higher polarization.

This paper is organized as follows. In section II, we discuss the geometry of the system and its connection to the one dimensional Hubbard model. In section III, we briefly discuss our finite-temperature TBA calculation scheme. In section IV, we discuss the compressibility calculations and their connections to the phase separation and re-entrant transition into a mixed phase. We devote section V to discussion of the experimental connections and we provide an experimental scheme to detect the phase separation. Finally in section VI, we draw our conclusions.

\section{II. The Model: A One dimensional Optical lattice and the Hubbard Model}

In general, a one dimensional optical lattice refers to an optical lattice generated by one set of laser standing waves. The result of combined trapping and a periodic potential gives a pancake-like shape of the surfaces of constant potential. However, the geometry we consider here is generated by a three-dimensional optical lattice where reduced dimensionality is achieved by freezing the atomic motion in the transverse direction. This can be done by operating two standing waves out of three mutually perpendicular laser standing waves at higher beam intensities. The higher intensities suppress tunneling in the transverse direction and create an array of one-dimensional lattice tubes. The dynamics of the atoms in one lattice tube can be modeled by the one dimensional Hubbard model given by,

\begin{eqnarray}
H = - t\sum_{<ij>,\sigma} c^\dagger_{i\sigma}c_{j\sigma}+U\sum_in_{i\uparrow}n_{i\downarrow} \\ \nonumber
- \mu \sum_{i\sigma} c^\dagger_{i\sigma}c_{i\sigma}-h\sum_{i\sigma} \sigma c^\dagger_{i\sigma}c_{i\sigma} \label{H1}.
\end{eqnarray}

\noindent The first term is the kinetic energy and is proportional to the tunneling amplitude $t$ between lattice sites $i$ and $j = i+1$. The operator $c^\dagger_{i\sigma} (c_{i\sigma})$ creates(destroys) a Fermi atom with hyperfine state denoted by pseudo-spin  $\sigma = \uparrow, \downarrow (\pm 1)$ at lattice site $i$. The second term describes the on-site interaction energy $U$. The density operator or the occupation number operator is $n_{i \sigma}= c^\dagger_{i\sigma} c_{i\sigma}$. Notice that $<ij>$ indicates only the nearest neighbor pair of sites and we neglect tunneling beyond the nearest neighbors. The average chemical potential $\mu =(\mu_\uparrow +\mu_\downarrow)/2$ and the chemical potential difference $h = (\mu_\uparrow - \mu_\downarrow)/2$, where $\mu_\sigma$ is the chemical potential of hyperfine state $\sigma$. Here we neglect the confinement harmonic trapping potential and consider the lattice tubes are homogenous in space. The effect of trapping potential is discussed in section V.

The tunneling amplitude and on-site interaction are related to the complete set of Wannier functions $w_{n,i}(\vec{r}) = \prod_{\alpha} w_n(\alpha-\alpha_i)$ localized at position $\vec{r}_i$ with band index $n$, where $\alpha = x, y, z$ are the components of Cartesian coordinates. As the band gap becomes larger than $U$ and temperature $T$, only the lowest band $n=0$ is populated. For deep lattices, the lattice potential at site $i$ can be approximated as a three-dimensional harmonic potential with frequency $\omega_\alpha = 2E_R\sqrt{s_\alpha}/\hbar$, where $s_\alpha E_R$ is the laser intensity of the standing wave in the $\alpha$ direction. The recoil energy $E_R = (\hbar k)^2/2m$ is the kinetic energy of an atom with mass $m$ and the momentum $\hbar k$ of a single lattice photon. For deep lattices, taking $w_0(\alpha-\alpha_i)$ as a ground-state harmonic oscillator function with frequency $\omega_\alpha$, the tunneling amplitude in the one dimensional geometry becomes $t = \int dx w^\ast_0(x-x_i)[-\frac{\hbar^2}{2m}\frac{\partial^2}{\partial x^2}+ V_0(x)]w_0(x-x_j)$. This is obtained from the Mathieu equation as $t = 4/\sqrt{\pi} E_Rs_x^{3/4}\exp{(-2\sqrt{s_x})}$, where $V_0(x) = s_x E_R\sin^2(kx)$ is the periodic potential generated by counter propagating lasers in the $x$ direction. The lattice constant $d = \lambda/2$ is related to the laser wave length $\lambda$, hence the wave vector $k = 2\pi/\lambda$. The on-site interaction $U = 4\pi\hbar^2a_s \int dx |w_0(x)|^4/m \propto a_s \sqrt{s_x}$. Notice that the on-site interaction $U$ can be repulsive or attractive depending on the free-space $s$-wave scattering length $a_s$. In the present work, we consider a tight one dimensional geometry in the $x$ direction with a positive $U$ modeled by Eq. 1. Notice that the ratio $t/U$ can easily be controlled by the laser intensity $I \propto s_x$ of the counter propagating lasers in the x-direction. In our model, the laser intensities in the transverse directions that are proportional to $s_y$ and $s_z$ are maintained at higher intensities so that the tunneling in the transverse direction is neglected.

\section{III. Thermodynamic Bethe ansatz method}

Lieb and Wu have shown that the model presented in the previous section is exactly solvable in one dimension using the thermodynamic Bethe-ansatz method~\cite{LW}. Following Takahashi~\cite{takahashi, essler}, the thermodynamic potential per site is given by

\begin{widetext}

\begin{eqnarray}
\Omega = e_0-\mu -k_BT \biggr\{\int^\pi_{-\pi} \rho_0(k) \ln[1+\xi(k)] dk + \int^\pi_{-\pi}\sigma_0(\Lambda) \ln[1+\eta_1(\Lambda)] d\Lambda \biggr\}.
\end{eqnarray}
\end{widetext}

\noindent The energy per site here is given as $e_0 = 2tI$, and two distribution functions of $k$'s and $\Lambda$'s are given by,

\begin{eqnarray}
\rho_0(k) &=& \frac{1}{2\pi} + \cos k\int^{\infty}_{-\infty}a_1(\Lambda-\sin k)\sigma_0(\Lambda) d\Lambda \\ \nonumber
\sigma_0(\Lambda) &=& \frac{1}{2\pi}\int^{\infty}_{-\infty}s(\Lambda-\sin k)dk.
\end{eqnarray}

\noindent The two additional expressions introduced in the equations are $a_1(x) = 4 u/[\pi (u^2 + 16 x^2)]$ and $s(x) = \csc (2\pi x/u)/u$ with $u = U/t$. The quantity $I$ is related to the $m$th order Bessel functions $J_m(x)$ through,

\begin{eqnarray}
I = -2\int_0^\infty\frac{J_0(x)J_1(x)}{x(1 + e^{ux/2})}dx.
\end{eqnarray}

The particle-hole ratios of $k$ excitations and $\Lambda$ excitations, $\xi(k)$ and $\eta_1(k)$ are obtained by an infinite set of nonlinear integral equations:

\begin{widetext}
\begin{eqnarray}
\ln\xi(k) &=& \frac{\kappa_0(k)}{T} + \int_{-\infty}^\infty d\Lambda s(\Lambda-\sin k) \ln\biggr(\frac{1+\eta_1^\prime(\Lambda)}{1+\eta_1(\Lambda)}\biggr) \\
\ln\eta_1(\Lambda)&=& s^*\ln[1+\eta_2(\Lambda)]-\int^\pi_{-\pi} s(\Lambda-\sin k) \ln[1+\xi^-1(k)]\cos k dk \\ \nonumber
\ln\eta_1^\prime(\Lambda)&=& s^*\ln[1+\eta_2^\prime(\Lambda)]-\int^\pi_{-\pi} s(\Lambda-\sin k)\ln[1+\xi(k)]\cos k dk \\ \nonumber
\end{eqnarray}

\noindent and for $j \geq 2$,

\begin{eqnarray}
\ln\eta_j(\Lambda)&=& s^*\ln\{[1+\eta_{j-1}(\Lambda)][1+\eta_{j+1}(\Lambda)]\} \\
\ln\eta_j^\prime(\Lambda)&=& s^*\ln\{[1+\eta_{j-1}^\prime(\Lambda)][1+\eta_{j+1}^\prime(\Lambda)]\}. \\ \nonumber
\end{eqnarray}
\end{widetext}

\noindent Here we use two integral functions given by $s^*f(\Lambda)\equiv\int^\infty_{-\infty}s(\Lambda-\Lambda^\prime)f(\Lambda^\prime)d\Lambda^\prime$ and $\kappa_0(k) \equiv -2t\cos k-4t\int^\infty_{-\infty}d\Lambda s(\Lambda-\sin k) \times Re\sqrt{1-(\Lambda-u i/4)^2}$. The average chemical potential and the chemical potential difference are entered in the formalism through the grand potential $\Omega$,

\begin{eqnarray}
\lim_{n\to\infty} \frac{\ln \eta_n(\Lambda)}{n} = \frac{2h}{T},
\end{eqnarray}

\noindent and
\begin{eqnarray}
\lim_{n\to\infty} \frac{\ln \eta_n^\prime(\Lambda)}{n} = \frac{U-2 \mu}{T}.
\end{eqnarray}

In order to calculate the thermodynamic potential numerically, one has to cut off the set of infinite equations at a finite number $j$. We achieve this by following the numerical procedure proposed by Takahashi \emph{et al}.~\cite{takahashiN}. The infinite set of equations is truncated by replacing $s(\Lambda)$ by $\delta(\Lambda)/2$ at $j > n_c$. Then the integral equations are converted into a set of matrix equations in which $2n_c +1$ unknown functions are represented in terms of discrete points of $k$ and $\Lambda$. These non-linear matrix equations are then solved iteratively using Newton's method for a given temperature ($T$), average chemical potential ($\mu$), and chemical potential difference ($h$). The details of the numerical procedure can be found in Refs.~\cite{takahashiN, theja1, theja2}. From the numerical solutions of the non-linear integral equations, we first calculate the thermodynamic potential $\Omega$ using Eq. (2), and then the particle density $n \equiv n_\uparrow + n_\downarrow = -\partial \Omega/\partial \mu$ and the magnetization (the density difference of two hyperfine states, $n_\uparrow - n_\downarrow$) $m = \partial \Omega/\partial h$ follow. The compressibility is then calculated numerically at a constant polarization $P = m/n$ using the second derivative of the thermodynamic potential with respect to the chemical potential~\cite{takahashiN, TU}.

\section{IV. The results: Identifying phase separation and re-entrant transition through Compressibility}

We examine the stability of the mixed phase through the sign of compressibility. Negative compressibility indicates an instability of the mixed phase, where the system enters into a phase-separated state.

\begin{figure}
\includegraphics[width=\columnwidth]{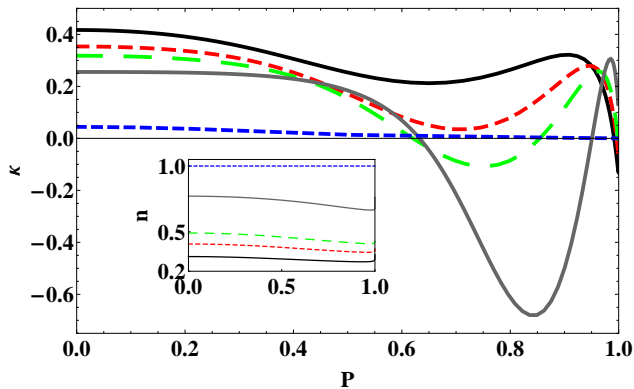}
\caption{(color online) Compressibility of a one dimensional lattice fermion system at interaction strength $U = 2 t$ and temperature $\beta = 30/t$. As shown in inset, the density is approximately fixed. The negative compressibility indicates the instability of the mixed phase against the phase-separated state. Notice that the plotted quantities are dimensionless.}\label{comp1}
\end{figure}

\noindent Figure~\ref{comp1} shows the compressibility at a constant interaction strength and a constant finite temperature for various values of density. The compressibility is always positive close to the densities of half filling and zero filling. However, away from these two limits, the compressibility becomes negative and then gets positive again as one increases the polarization. Notice the compressibility for filling factors $ n \simeq 0.5$ and $n \simeq 0.75$ in the figure. The negative compressibility indicates the instability of the mixed phase, meaning that the system is phase separated into two different distinct phases corresponding to their pseudo-spins. A further increase of polarization causes the system to make a re-entrant transition into the mixed phase. In the mixed phase, both spin components coexist in the same region of space. This positive compressibility at higher polarization itself does not guarantee the stability of the mixed phase over the phase separated phase. One has to compare the energies of phase-separated state and the mixed phase to determine the stability. The zero-temperature stability of the mixed phase at higher polarizations is justified in Ref.~\cite{LB}. This justification has been confirmed by comparing the ground-state energies in both mixed and phase separated states using both weak and strong coupling approaches. We believe this is true even for finite temperatures. The comparison of finite temperature energies of the phase-separated state and the mixed state is not trivial and these calculations are beyond the scope of the present paper.

\begin{figure}
\includegraphics[width=\columnwidth]{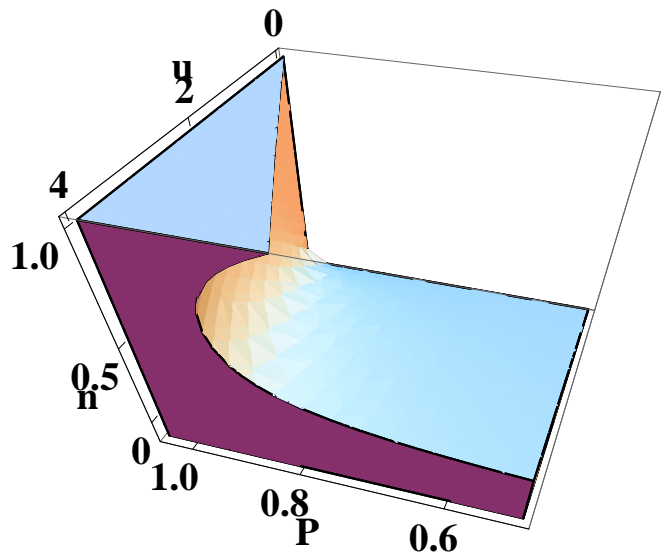}
\caption{(color online) Zero temperature phase diagram of one dimensional lattice fermions in polarization ($P$), on-site interaction ($U$), and density ($n$) parameter space. The phase diagram is constructed from bosonization Theory. While the shaded region represents the phase separated state, the unshaded region represents the mixed state.}\label{ZeroT}
\end{figure}

The zero-temperature instability of the mixed phase at higher polarizations has already been established within the bosonization theoretical frame work~\cite{bosT}. Bosonization theory suggests that phase separation occurs for $U/t \geq 4 \pi \{\sin[\pi(n+m)/2]\sin[\pi(n-m)/2]\}^{1/2}$. As shown in Fig~\ref{ZeroT}, the mixed phase is stable only at higher densities, low polarizations, and low interaction strengths. The phase-separated state is stable at higher polarizations; however, unlike finite temperatures, the system does not make a re-entrant transition into the mixed phase at zero temperature.

\begin{figure}
\includegraphics[width=\columnwidth]{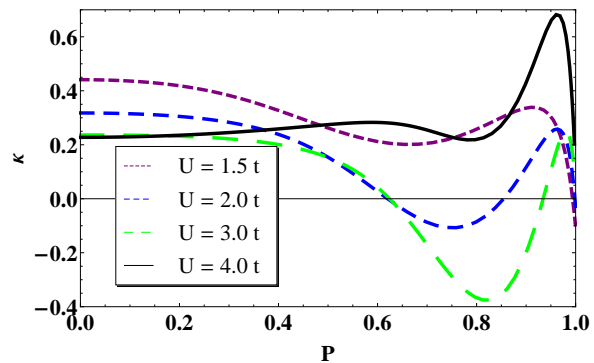}
\caption{(color online) Compressibility of a one dimensional lattice fermion system at a constant temperature $\beta = 30/t$ for various interaction strengths. See FIG.~\ref{comp1} caption for details.}\label{compFTFM}
\end{figure}

\begin{figure}
\includegraphics[width=\columnwidth]{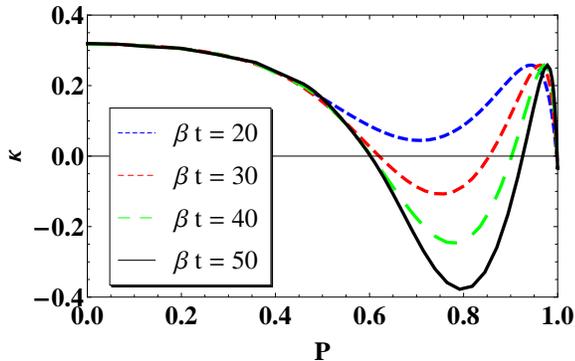}
\caption{(color online) Compressibility of a one dimensional lattice fermion system at a constant interaction strength  $U = 2 t$ for different temperatures. See FIG.~\ref{comp1} caption for details.}\label{compFUFM}
\end{figure}

The compressibility at various interaction strengths and temperatures is shown in Fig.~\ref{compFTFM} and Fig.~\ref{compFUFM} respectively. Here the compressibility is calculated at a desired polarization $P$ by varying the chemical potential difference $h$ and keeping the average chemical potential at a representative fixed value, $\mu = 3 t$. Notice that the mixed phase is stable for the entire range of polarization at smaller and larger interactions. This can be justified by the compressibility at the infinite interaction and non-interacting limits. At the infinite interaction limit, the TBA equations can be solved to get analytical results~\cite{InU}. In the limit $U \rightarrow \infty$, the polarization $P = \tanh(\beta h)/2$ and the compressibility can be calculated as $\kappa = 2 \beta \cosh(\beta h)/4 \int_0^{\pi}f(k) dk$, where $f(k) = \exp[\beta(\mu+2t \cos k)]/(1+\cosh(\beta h)\exp[\beta(\mu+2t \cos k)])^2$. As $f(k) > 0 $ for all $k$ values, the compressibility at the infinite interaction limit is always positive. This is intuitive as the system can be considered as spinless fermions in this limit. On the other hand, in the limit $U \rightarrow 0$, no phase separation occurs as the system consists of non interacting fermions. Again, the negative compressibility at intermediate interactions suggests the phase separation into two pseudo spin states.

 Consider the temperature dependence shown in FIG.~\ref{compFUFM}. The mixed phase makes a transition into a phase separated state and then makes a re-entrant transition into mixed phase at higher polarization for low temperatures. In contrast, the mixed phase is stable for higher temperatures (small $\beta$) over the entire range of polarization. The high temperature expansion of the thermodynamic potential for the one dimensional Hubbard model has been carried up to the fourth order in $\beta$ by Charret \emph{et al}~\cite{cha} and up to the sixth order in $\beta$ by Takahashi \emph{et al}~\cite{takahashiN}. By using the 6th order expansion, we confirm the positive compressibility at higher temperatures by an analytic calculation. The high temperature expansion of the compressibility and the polarization up to the sixth order is given in the Appendix.

Notice that compressibility is a non-monotonic function of polarization for all temperatures and interactions. In contrast, compressibility is a non-monotonic function of the interaction parameter only for larger polarizations. However, as evident from the Fig.~\ref{compFTFM}, compressibility is a monotonic function of temperature for the entire range of polarizations.

It is worth mentioning that a small density imbalance can be induced in condensed matter electronic systems by applying an external magnetic field. Thermodynamic properties of such one dimensional systems are thoroughly discussed in Ref.~\cite{essler}. Though finite temperature compressibility as a function of polarization is not discussed in there, special attention has been given to the ground-state properties such as susceptibility, magnetization, and densities~\cite{note}.

\section{V. CONNECTIONS TO EXPERIMENTS}

Recent progress in experimental techniques with ultra-cold atoms, such as single-site detection ~\cite{jacob, bakr}, noise correlations~\cite{alt, foll}, Bragg scattering~\cite{we}, and \emph{in situ} imaging in the lattice scaling~\cite{bakr}, allows one to probe the density variations in cold-atom experiments. For the case of equal-population two-component fermions on a three-dimensional cubic lattice, the compressibility has already been measured~\cite{np}.

\begin{figure}
\includegraphics[width=\columnwidth]{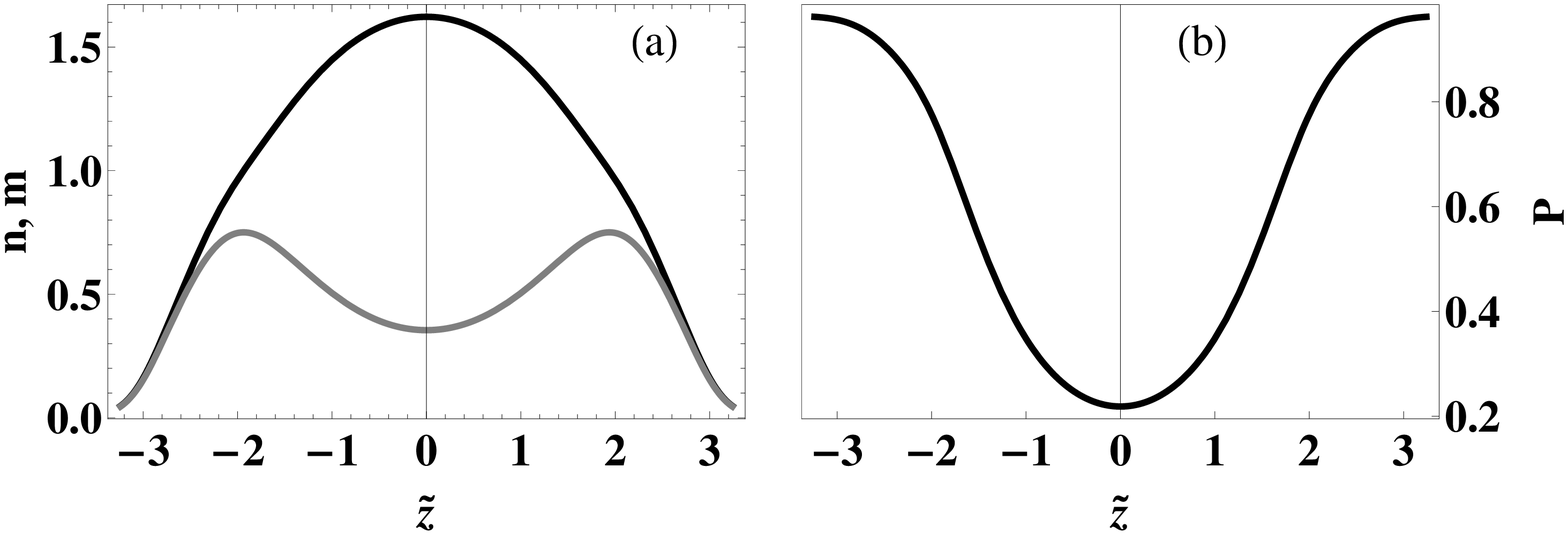}
\caption{Spatial variations of the atom density $n(z) = n_\uparrow(z) + n_\downarrow(z)$ [black curve in panel (a)], magnetization $m(z) = n_\uparrow(z) - n_\downarrow(z)$ [gray curve in panel (a)]) , and polarization $P(z) = [n_\uparrow(z) - n_\downarrow(z)]/[n_\uparrow(z) + n_\downarrow(z)]$ [panel (b)]. We define the scaled length $\tilde{z} = z\sqrt{m\omega^2/2}$, where $\omega$ is the one dimensional trapping frequency. We fixed the on-site interaction ($U =2 t$) and the inverse temperature ($\beta t= 1$). }\label{pol}
\end{figure}

Though we neglected it in this study, the underlying harmonic trapping potential
present in all cold gas experiments causes the density to vary across the lattice. See FIG.~\ref{pol}. By combining the TBA solutions with the local density approximation (LDA), we then extract the local density $n(z)$, magnetization $m(z)$, and then polarization $P(z)$. In LDA, the external trapping potential $V_{i} =
m\omega^2 z^2/2$ at site $i$ is related to the local chemical potential through the relation $\mu_i = \mu_0 - V_i$, where $\omega$ is the one-dimensional trapping potential, $\mu_0$ is the central chemical potential and $z = id$, with lattice constant $d$ the spatial coordinate. As shown in Fig.~\ref{pol}, the density monotonically decreases, while polarization monotonically increases from the center to the edge of the trap~\cite{theja2}. This trapping potential induced inhomogeneity allows both mixed-phase and phase-separated states to exist simultaneously inside the trap. At the center of the trap, the density is higher and the polarization is lower. On the other hand, the polarization is higher and the density is lower at the edge of the trap. As a result, the mixed-phase should exist at the center and at the edge of the trap. However, depending on the density, the phase separated state can exists in the middle (not the center) of the trap. Therefore, by adjusting the total density in the trap, any polarization induced phase separation and re-entrant transition can be investigated experimentally with currently available experimental techniques.

\section{VII. CONCLUSIONS}

In conclusion, we considered two-component Fermi atoms in a highly tunable optical lattice to study the phase separation of fermions in one dimension. We have calculated the compressibility of one dimensional lattice fermions using the thermodynamic Bethe ansatz method. We find that compressibility is a non-monotonic function of polarization. At filling factor $0 < n <1$, with low temperatures and intermediate interactions, compressibility becomes negative, indicating instability of the mixed-phase state towards the phase separated state. For some parameters at higher polarizations, compressibility becomes positive again indicating a re-entrant transition in to a mixed phase. These phase-separation and re-entrant transitions can be detected by using currently available experimental techniques.

\section{VIII. Appendix: High Temperature expansion of the compressibility}

In this appendix, we provide compressibility and polarization up to the sixth order in $\beta$. Using the high-temperature expansion of the thermodynamic potential~\cite{cha, takahashiN}, we find the compressibility,

\begin{widetext}

\begin{eqnarray}
\kappa = \frac{\beta}{2} - \frac{U \beta^2}{8} + \frac{\kappa_3}{32} \beta^3 + \frac{\kappa_4}{384} \beta^4 + \frac{\kappa_5}{1536} \beta^5 + \frac{\kappa_6}{30720} \beta^6 + \mathcal{O}(\beta^7),
\end{eqnarray}

\noindent where the coefficients at each order are given by

\begin{eqnarray}
\kappa_3 = -4 h^2 - 8 t^2 - (U - 2 \mu)^2,
\end{eqnarray}

\begin{eqnarray}
\kappa_4 = U (48 t^2 + 13 U^2 - 48 U \mu + 48 \mu^2),
\end{eqnarray}

\begin{eqnarray}
\kappa_5 = 32 h^4 + 192 t^4 +
 96 t^2 (U - 2 \mu)^2 - (U - 2 \mu)^2 (7 U^2 + 8 U \mu -
    8 \mu^2) +
 4 h^2 [15 U^2 + 2 t^2 (48 + U^2) - 48 U \mu + 48 \mu^2],
\end{eqnarray}

\noindent and
\begin{eqnarray}
\kappa_6 = -U [-240 h^4 + 2880 t^4 + 107 U^4 + 360 h^2 (U - 2 \mu)^2 -
   760 U^3 \mu + 2120 U^2 \mu^2 - 2720 U \mu^3 + 1360 \mu^4  \\ \nonumber +
   240 t^2 (11 U^2 - 40 U \mu + 40 \mu^2)].
\end{eqnarray}

The sixth-order expansion of polarization becomes,

\begin{eqnarray}
P = \frac{h\beta}{2} - \frac{h(U-\mu) \beta^2}{4} + \frac{P_3}{24} \beta^3 + \frac{P_4}{48} \beta^4 + \frac{P_5}{3840} \beta^5 + \frac{P_6}{46080} \beta^6 + \mathcal{O}(\beta^7),
\end{eqnarray}

\noindent where the coefficients of higher orders are,

\begin{eqnarray}
P_3 = -h (h^2 + 6 t^2),
\end{eqnarray}

\begin{eqnarray}
P_4 = -h [4 h^2 + 12 t^2 + (U - \mu)^2] (U - \mu)
\end{eqnarray}

\begin{eqnarray}
P_5 = h \{16 h^4 + 5 t^2 [96 t^2 + (12 + U^2) (U - 2 \mu)^2] +
   40 h^2 [8 t^2 - 3 (U - \mu)^2)]\},
\end{eqnarray}

\noindent and

\begin{eqnarray}
P_6 = h \{816 h^4 (U - \mu) +
   20 (h^2) [
     16 (U - \mu)^3 +
      3 t^2 (160 U + U^3 - 160 \mu - 2 U^2 \mu)] +
   3 [2880 t^4 (U - \mu) + 32 (U - \mu)^5 \\ \nonumber +
      5 t^2 (U^5 - 6 U^4 \mu + 576 U \mu^2 - 256 \mu^3 -
         8 U^2 \mu (60 + \mu^2) + 4 U^3 (38 + 3 \mu^2))]\}.
\end{eqnarray}

\end{widetext}

\section{V. Acknowledgements}

We thank Joseph Newton for critical comments on the manuscript.

\end{document}